\documentclass[journal=jpcafh,manuscript=article]{achemso}

\usepackage{graphicx}
\usepackage{color}
\usepackage{dcolumn}
\usepackage{multirow}
\usepackage{bm}
\usepackage[normalem]{ulem}

\usepackage{cmbright}
\usepackage{float}
\usepackage{siunitx}
\usepackage[section]{placeins}
\usepackage{booktabs}
\usepackage[version=4]{mhchem}
\usepackage{xcolor}
\usepackage{tablefootnote}
\usepackage{colortbl}
\usepackage{adjustbox}
\SectionNumbersOn

\newcommand{\ciab}[1]{\text{CI}^{\text{S} _\text{0}/\text{S}_\text{1}}_{#1}}

\title{A Theoretical Perspective on the Photochemistry of Boron--Nitrogen Lewis Adducts}

\author{Emanuele Marsili}
\author{Basile F. E. Curchod}
\email{basile.curchod@bristol.ac.uk}
\affiliation[University of Bristol]{Centre for Computational Chemistry, School of Chemistry, University of Bristol, Bristol BS8 1TS, UK}

\begin{document}

\begin{abstract}

Boron--Nitrogen (\ce{B-N}) Lewis adducts form a versatile family of compounds with numerous applications in functional molecules. Despite the growing interest in this family of compounds for optoelectronic applications, little is currently known about their photophysics and photochemistry. Even the electronic absorption spectrum of ammonia borane, the textbook example of a \ce{B-N} Lewis adduct, is unavailable. Given the versatility of the light-induced processes exhibited by these molecules, we propose in this work a detailed theoretical study of the photochemistry and photophysics of simple \ce{B-N} Lewis adducts. We used advanced techniques in computational photochemistry to identify and characterize the possible photochemical pathways followed by ammonia borane, and extended this knowledge to the substituted \ce{B-N} Lewis adducts pyridine-borane and pyridine-boric acid. The photochemistry observed for this series of molecules allows us to extract qualitative rules to rationalize the light-induced behavior of more complex \ce{B-N} containing molecules.

\end{abstract}

\begin{tocentry} 
\includegraphics[width=1.0\textwidth]{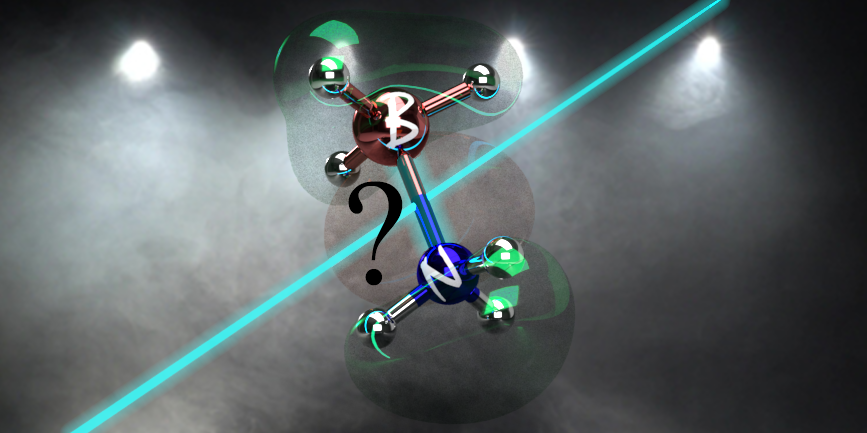}
\end{tocentry}


\section{Introduction}
The versatility of Lewis acid-base adducts, characterized by a dative bond between a Lewis acid and Lewis base, has been extensively used for a plethora of molecular applications like organic synthesis,\cite{mccahill2007reactivity, dureen2009terminal} catalysis,\cite{xu2014probing, ghuman2016photoexcited, li2020co2} production of new types of dyes for solar cells\cite{nugegoda2022lewis} or polymers.\cite{adelizzi2020long, yolsal2021cyclic} More recently, a particular boost of interest has emerged for the photochemistry and photophysics of Lewis adducts, with potential applications for functional materials,\cite{sun2019bn, adelizzi2020long, shi2022dynamic} sensors,\cite{osumi2016boron, hou2018stimuli} or optoelectronic devices.\cite{matsuo2014photodissociation, ando2021boron, vanga2022linear} For example, the inclusion of a boron atom in polycyclic aromatic hydrocarbons (PAH) leads to highly stable photoactive molecules, whose optical properties can be altered upon adduct formation with Lewis bases to act as sensors.\cite{matsuo2014photodissociation, ando2021boron} These \ce{B-N} Lewis adducts, however, exhibit a rather peculiar photochemical behavior involving excited-state photodissociation and an unexpected double fluorescence. Interestingly, the light-induced dissociation is not unique to this class of Lewis adducts and has been observed for several compounds containing a constrained boron center paired with a relatively weak Lewis base.\cite{kano2005photoswitching, matsumoto2015synthesis, matsumoto2017design, hou2018stimuli, shi2022dynamic}

Very little is known to date about the photochemistry and photophysics of \ce{B-N} Lewis adducts. This observation comes as a surprise considering the important body of work on this class of molecules for functional materials and their rather unexpected photochemical properties. Only a few studies tried to connect the strength of the \ce{B-N} bond to the observed photodissociation behavior.\cite{matsuo2014photodissociation, matsumoto2015synthesis, matsumoto2017design, ando2021boron} However, the relation between the strength of the \ce{B-N} bond in the ground electronic state of the molecule and its potential weakening in the excited states is far from trivial given our general lack of understanding of the electronic-state characters involved in the photochemistry for these molecules. 

Perhaps even more surprising is the fact that our unawareness of the photochemistry and photophysics of Lewis adduct extends to ammonia borane, \ce{H3N-BH3}, the simplest \ce{B-N} Lewis adduct. This molecule, often considered as a prototypical model to discuss \ce{B-N} adducts, has received increased attention in the past years for its connection with hydrogen-storage material.\cite{denney2006efficient, zhang2017ruthenium, al2013mechanistic, smythe2010ammonia, bhattacharya2014mechanistic} As a result, the physical properties of \ce{H3N-BH3} were carefully analyzed by several spectroscopic techniques such as microwave,\cite{thorne1983microwave} NMR,\cite{penner1999deuterium} IR,\cite{goubeau1961borinhydrazin} Raman,\cite{taylor1958vibrational} single-photon ionization,\cite{yuan2016dynamics} photoelectron spectroscopy,\cite{lloyd1972photoelectron, schleier2022ammonia} X-ray crystallography,\cite{hoon1983molecular} neutron powder diffraction,\cite{hess2009neutron} and inelastic neutron scattering.\cite{allis2004inelastic} In addition, numerous computational works focused on obtaining ground-state equilibrium geometries,\cite{binkley1983theoretical, binkley1983theoretical, lee2009comparative} vibrational frequencies and zero-point energy,\cite{binkley1983theoretical, sams2012vapor, lee2009comparative}, as well as its charge transfer characteristics\cite{mo2004charge} and ground-state dissociation barrier.\cite{zimmerman2011dynamic} Besides a single theoretical absorption cross-section of \ce{H3N-BH3},\cite{naganathappa2015mono} no detailed information about the photochemistry and photophysics of the prototypical ammonia borane appear to exist in the literature, to the best of the authors' knowledge.

\begin{figure}[h]
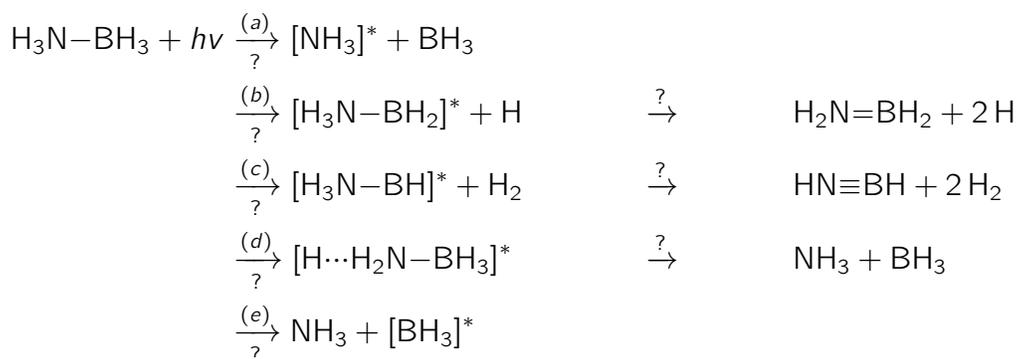

    \centering
    \begin{align*}
\ce{    H3N-BH3 + $hv$} &\xrightarrow[?]{(a)} \ce{[NH3]^* + BH3} \\
                        &\xrightarrow[?]{(b)} \ce{[H3N-BH2]^* + H}  && \xrightarrow{?}   && \ce{H2N=BH2 + 2 H} \hspace{0.1cm} \\
                        &\xrightarrow[?]{(c)} \ce{[H3N-BH]^* + H2}  &&\xrightarrow{?}    && \ce{HN#BH + 2 H2} \hspace{0.1cm} \\
                        &\xrightarrow[?]{(d)} \ce{[H\bond{...}H2N-BH3]^*} &&\xrightarrow{?} && \ce{NH3 + BH3}    \\
                        &\xrightarrow[?]{(e)} \ce{NH3 + [BH3]^*} \\
    \end{align*}
    \caption{Possible photodissociation channels for \ce{H3N-BH3}. \ce{[A]^*} denotes that the species \ce{A} is in an excited electronic state. Question marks are added to stress that these mechanisms are purely speculative -- it is the goal of this work to assess whether such processes are possible.}
    \label{fig: scheme photodissociation}
\end{figure}

Inspired by the VUV photodissociation of ethane and foreshadowing some of the results presented in this work,\cite{chang2020ultraviolet} we summarize in Fig.~\ref{fig: scheme photodissociation} the potential photodissociation channels that \ce{H3N-BH3} may encounter. Upon light excitation, ammonia borane can relax to its ground electronic state via internal conversion. The hot ammonia borane formed in the ground electronic state could then possible dissociate into \ce{NH3} and \ce{BH3}, with both fragments being in their ground electronic state (Fig.~\ref{fig: scheme photodissociation}d). Alternatively, \ce{H3N-BH3} could dissociate in an excited electronic state yielding \ce{NH3} or \ce{BH3}, with one of these fragments still being in an excited electronic state (Fig.~\ref{fig: scheme photodissociation}a and Fig.~\ref{fig: scheme photodissociation}e). Photoexcitation of \ce{H3N-BH3} could also produce either \ce{H2} and \ce{H2N=BH2} (Fig.~\ref{fig: scheme photodissociation}b) or \ce{2H2} and \ce{HN#BH} (Fig.~\ref{fig: scheme photodissociation}c). The observation of such potential nonradiative deactivation channels is likely to depend on the wavelength of the light source used to photoexcite ammonia borane. One may also wonder how altering the electronic properties of the Lewis base or the Lewis acid forming the adduct could alter the nonradiative deactivation pathways of a \ce{B-N} adduct. In the following, we will focus on simple alterations of ammonia borane such as pyridine-borane (\ce{Py-BH3}) or pyridine-boric acid (\ce{Py-B(OH)3}).

Hence, the main objectives of this work consist in (i) investigating the excited electronic states of \ce{H3N-BH3}, (ii) identifying the possible photodissociation pathways of \ce{H3N-BH3} and their dependence on the excitation wavelength, (iii) extending this knowledge to the more complex \ce{B-N} Lewis adducts \ce{Py-BH3} and \ce{Py-B(OH)3}, and (iv) proposing qualitative guidelines to predict the photochemistry of general \ce{B-N} adducts. To achieve these goals, we propose to investigate theoretically these three \ce{B-N} Lewis adducts by exploiting a combination of quantum-chemical calculations and excited-state (nonadiabatic) molecular dynamics simulations. These computational techniques give access to potential energy curves describing the possible nonradiative pathways, photoabsorption cross-sections, and wavelength-dependent quantum yields (building upon a protocol developed in our group to study atmospheric volatile organic compounds\cite{prlj2020theoretical, marsili2022theoretical, hutton2022photodynamics, prlj2021calculating}). We start this work by detailing the computational methodologies employed for our calculations (Sec.~\ref{sec: computational details}), before presenting our main results in Sec.~\ref{sec: results and discussion}. We first focus on the possible photochemical pathways experienced by ammonia borane (Sec.~\ref{subsec: ammonia-borane}), and offer a benchmark for the level of electronic-structure theory used in this work. We then present our results for the photochemistry of \ce{Py-BH3}  (Sec.~\ref{subsec: pyridine-borane}) and \ce{Py-B(OH)3} (Sec.~\ref{subsec: pyridine-boric acid}) and contrast them with the excited-state dynamics of ammonia borane. Finally, we determine a set of qualitative rules for the photochemistry of \ce{B-N} adducts and used them to interpret experimental findings from the literature (Sec.~\ref{sec: implications of the Lewis adduct photodissociation}).  

\section{Computational details}
\label{sec: computational details}

\subsection{Electronic structure and benchmark of electronic-structure methods for \ce{H3N-BH3} }
\label{subsec: electronic structure}
Given the potentially rich photochemistry of \ce{B-N} Lewis adducts, an adequate choice of electronic-structure methods is critical to correctly capture the different photodissociation pathways. Following an extensive benchmark (see below), we opted for a combination of MP2 (M\o{}ller-Plesset perturbation theory up to second-order) and ADC(2) (algebraic diagrammatic construction of second order).\cite{hattig2005structure, dreuw2015algebraic} All calculations were performed with Turbomole 7.4.1,\cite{furche2014turbomole} using frozen cores and the resolution of identity (RI) approximation.\cite{weigend2002efficient} 

Results obtained with MP2 and ADC(2) at the minimum-energy structure in S$_0$ were validated against EOM-CCSD (equation-of-motion coupled-cluster singles and doubles, performed with Gaussian09\cite{frisch2009gaussian}), the multiconfigurational method SA-CASSCF (state-averaged complete active space self-consistent field),\cite{roos1980complete, roos1980complete1, siegbahn1980comparison} and XMS-CASPT2 (extended multi-state complete active space second-order perturbation theory ).\cite{shiozaki2011communication, park2017fly} Different basis sets were tested: the correlation-consistent polarized (cc-pVDZ, cc-pVTZ, cc-pVQZ, and the respective augmented sets)\cite{woon1993gaussian} and the Karlsruhe (def2-SVPD) basis sets.\cite{weigend2005balanced, weigend2006accurate, rappoport2010property} BAGEL 1.2 was used for all SA-CASSCF and XMS-CASPT2 calculations.\cite{shiozaki2018bagel} SA-CASSCF and XMS-CASPT2 calculations were also employed to assess the reliability of ADC(2) along the photodissociation pathways of \ce{H3N-BH3}, particularly focusing on the ground electronic state and its potential multireference character. All calculations using BAGEL employed density fitting (DF) and frozen cores. For all XMS-CASPT2 calculations, we employed a real vertical shift of 0.5 Hartree and the SS-SR contraction scheme. 

For the benchmark of the electronic-structure methods, we investigated four potential dissociation channels for \ce{H3N-BH3}, namely (a) the \ce{B-N} dissociation, (b) the \ce{B-H} dissociation, (c) the concerted dissociation of \ce{H2} via two \ce{B-H} bonds breaking, and (d) the \ce{N-H} elongation. These processes were studied via a combination of relaxed scans in the ground and/or first excited electronic state and linear interpolation in internal coordinates (LIICs) connecting critical points of the potential energy surfaces (PESs). Relaxed scans in the ground state -- used for (a) and (b) -- and in the first excited state -- used for (d) -- have been performed with MP2/aug-cc-pVDZ and ADC(2)/aug-cc-pVDZ, respectively. LIICs, connecting the Franck-Condon (FC) point (optimized with MP2/aug-cc-pVDZ), the S$_1$ minimum (optimized with ADC(2)/aug-cc-pVDZ), and the $\ciab{\text{BH}}$ and $\ciab{\text{H}_\text{2}}$ minimum-energy conical intersections (MECIs, optimized with XMS-CASPT2/aug-cc-pVDZ), were used to describe the dissociation channels (b) and (c). To benchmark the results obtained from MP2 and ADC(2), we obtained XMS-CASPT2/aug-cc-pVDZ energies along these relaxed scans and LIICs, using two different combinations of active spaces and state averaging. The first combination, used for pathways (a), used a state averaging over 13 singlet states and an active space composed of 8 electrons in 9 orbitals (see Fig.~S1 for a representation of the SA(13)-CASSCF(8/9) natural orbitals and their labeling). The second combination, used for pathways (b), (c), and (d), employed 5 singlet states for the state averaging and an active space of 4 electrons in 8 orbitals (see Fig.~S2 for a representation of the SA(5)-CASSCF(4/8) natural orbitals and their labeling). 

\subsection{Ground-state sampling and photoabsorption cross-sections}
\label{subsec: ground-state sampling and photoabsorption cross-sections}
For each molecule studied -- ammonia borane, pyridine-borane, and pyridine-boric acid -- a set of 500 initial conditions were sampled from an \textit{ab initio} molecular dynamics run using MP2/aug-cc-pVDZ for the electronic-structure theory and a quantum thermostat (QT), performed with the ABIN code\cite{hollas2021abin} coupled to the Turbomole 7.4.1\cite{furche2014turbomole} Parameters for the quantum thermostat were taken from the GLE4MD webpage,\cite{gle4md} using a target temperature T = 296 K and the following parameters: Ns = 6, $\hbar\omega/kT$ = 20, strong coupling. The time step for the \textit{ab initio} molecular dynamics was 20 atomic time units (atu). The equilibration time was determined by monitoring the convergence of the average kinetic energy temperature. The nuclear configurations, used subsequently as initial conditions with the corresponding nuclear velocities, were sampled each 2000 atu during the production run. 

The initial conditions were used to calculate the photoabsorption cross-sections of \ce{H3N-BH3}, \ce{Py-BH3}, and \ce{Py-B(OH)3} at the ADC(2)/aug-cc-pVDZ and ADC(2)/cc-pVDZ level of theory. None of the molecules studied contain a carbonyl group, which leads to artificial crossings between the first and the ground electronic states with ADC(2)/MP2.\cite{marsili2021caveat} An augmented basis set is needed to describe the Rydberg character of the \ce{H3N-BH3} excited states properly. In contrast, the use of an augmented basis is less important for the \ce{Py-BH3} and \ce{Py-B(OH)3} adducts, characterized by only low-lying valence excited states (see further discussion below). All spectral transitions were broadened with Lorentzians using a phenomenological broadening of 0.05 eV. The resulting photoabsorption cross-sections for \ce{H3N-BH3} and \ce{Py-BH3} were obtained by averaging the contribution of all 500 geometries using the nuclear ensemble approach (NEA).\cite{crespo2014spectrum} Due to the weak \ce{B-N} bond of the \ce{Py-B(OH)3} adduct, 97 out of the 500 initial conditions sampled from the QT dynamics had a \ce{B-N} bond longer than 2.1\AA. To avoid any bias in our result, we calculated the photoabsorption cross-section employing the 403 initial conditions exhibiting a \ce{B-N} bond shorter than 2.1\AA (see Fig.~S3 for the \ce{B-N} distribution and Fig.~S4 for the photoabsorption cross-section). 
The NEA and the spectrum were calculated with Newton-X version 2.4.\cite{barbatti2014newton}

\subsection{Excited-state dynamics and quantum yields}
\label{subsec: excited-state dynamics and quantum yields}
The excited-state (nonadiabatic) molecular dynamics simulations were performed with the trajectory surface hopping (TSH) algorithm.\cite{tully1990molecular} TSH dynamics were carried out using ADC(2)/aug-cc-pVDZ for \ce{H3N-BH3} and ADC(2)/cc-pVDZ for \ce{Py-BH3} and \ce{Py-B(OH)3}, with a time step of 0.5 fs and using Newton-X version 2.4 coupled with Turbomole.\cite{plasser2014surface} The nonadiabatic couplings were obtained using the overlap-based time-derivative couplings computed using the orbital derivative scheme,\cite{ryabinkin2015fast} and the kinetic energy was adjusted by rescaling the nuclear velocity vector isotropically following a successful hop. The electronic populations were corrected to prevent overcoherence using the energy-based decoherence correction of Granucci and Persico.\cite{granucci2007critical}

The TSH dynamics were started from the initial conditions used for the NEA (see above), grouped into different energy windows based on the calculated photoabsorption cross-sections of \ce{H3N-BH3}, \ce{Py-BH3}, and \ce{Py-B(OH)3}. Three energy windows with a width of 1.0 eV were selected for \ce{H3N-BH3} and \ce{Py-BH3}. For \ce{H3N-BH3}, the windows were centered at 6.0, 7.0, and 8.0 eV, while for \ce{Py-BH3} the three windows were centered at 4.0, 5.0, and 6.0 eV. Two windows instead were selected for \ce{Py-B(OH)3}, centered at 4.75 and 5.25 eV, each with a width of 0.5 eV. The choice of the width is somewhat arbitrary and basically guided by the balance between computational cost and the interplay between excited states observed from the photoabsorption cross-sections. The initial conditions were selected randomly in each window (\textit{i.e.}, not weighted by their individual oscillator strength), and their statistics are reported in Table~S1. For \ce{Py-B(OH)3}, we only used initial conditions with a \ce{B-N} bond shorter than 2.1\AA. All TSH trajectories were stopped when the S$_0$/S$_1$ energy gap became smaller than 0.05 eV and the last point of the dynamics is used to assess the type of nonradiative process suffered by the molecule -- we called this analysis the 'fraction of trajectories per CI' in the following. The standard deviations of the fraction of trajectories per CI (or photodissociation channel) were estimated following the method of Persico and Granucci.\cite{persico2014overview}

\section{Results and discussion}
\label{sec: results and discussion}

\subsection{Photochemistry of ammonia borane}
\label{subsec: ammonia-borane}

In the following, we focus our attention on the different deactivation pathways suffered by ammonia borane upon photoexcitation in its lowest excited electronic states. We first discuss the electronic character of the low-lying singlets states of \ce{H3N-BH3} in the FC region, before investigating the behavior of its electronic energies along a stretch of the \ce{B-N} bond. Inspired by the VUV photodissociation of ethane,\cite{chang2020ultraviolet} which is structurally and electronically similar to \ce{H3N-BH3}, we then investigate the potential H and \ce{H2} photodissociation channels of ammonia borane. In contrast with ethane,\cite{chang2020ultraviolet} the introduction of the B and N atoms in ammonia borane produces a strong asymmetry in the electronic structure of the excited electronic state, leading to a possible different behavior of the molecule when H photodissociation is triggered from the \ce{BH3} or the \ce{NH3} moiety. 

The low-lying excited electronic states of \ce{H3N-BH3} are characterized by a strong Rydberg character which imposes some constraints on the theoretical methods that can be used for their description. For example, it is well established that LR-TDDFT, within its practical approximations, usually yields a poor description of Rydberg states.\cite{zyubin2003performance} In addition, the high density of electronic states in the UV-VUV regime challenges the use of multireference methods based on a state-averaging procedure, in particular for excited-state dynamics. In this context, ADC(2) appears to be a viable alternative, given its compromise between efficiency and accuracy. In the following sections, we propose a discussion of the electronic states of ammonia borane and its possible photodissociation channels, combined with a careful benchmark of ADC(2) to assess its validity along the different decay pathways of this molecule.

\subsubsection{Vertical excitation energies of ammonia borane}
\label{subsec: vertical excitation energies of ammonia borane}
We first compare the excitation energies obtained with ADC(2), EOM-CCSD, and XMS-CASPT2 (using different basis sets) to confirm that ADC(2) can provide an adequate description of the Rydberg excited electronic states of \ce{H3N-BH3}. The combination of ADC(2) with the aug-cc-pVDZ basis set yields vertical transition energies and oscillator strengths in excellent agreement with EOM-CCSD/aug-cc-pVQZ for the lowest nine excited states investigated (see Table~S2). 
These results are further validated by XMS-CASPT2 calculations, even though some oscillator strengths appear weaker with this method -- a possible weakness of XMS-CASPT2 already highlighted in earlier work (see, for example, Ref.~\citenum{doi:10.1021/acs.jpca.3c02333}). 
Based on this first validation of ADC(2)/aug-cc-pVDZ in the FC region, we now extend our benchmark to the investigation of the photodissociation pathways expected for \ce{H3N-BH3}. All calculations presented from this point on will use an aug-cc-pVDZ basis set, except if stated otherwise.

\subsubsection{Pathway (a): photodissociation of the B--N bond in ammonia borane}
\label{subsubsec: NH3BH3 adduct photodissociation}
Let us first focus on the dissociation of the \ce{B-N} bond. In the ground electronic state, the \ce{B-N} dissociation follows a heterolytic cleavage such that the Lewis acid and base are released as neutral species. But how does such a dissociation take place in the excited electronic states? 

\begin{figure}[h]
    \centering
    \includegraphics[width=0.80\textwidth]{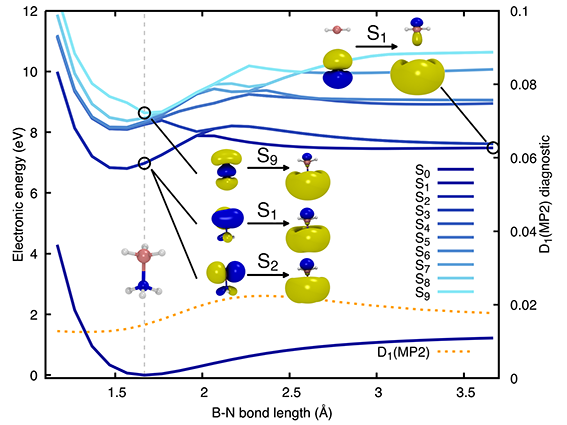}
    \caption{Relaxed scan along the \ce{B-N} bond length of \ce{H3N-BH3} obtained for the ground electronic state with MP2/aug-cc-pVDZ. Excited electronic energies were obtained with ADC(2)/aug-cc-pVDZ. The FC point is indicated by the gray vertical dashed line with the corresponding geometry as inset. The lowest nine electronic states are depicted in shades of blue -- from S$_0$ in dark blue to S$_9$ in light blue. The D$_1$ diagnostic for the MP2/aug-cc-pVDZ ground state is reported as a dotted orange line. The NTOs of the S$_1$/S$_2$ (E) and S$_9$ (A$_1$) electronic states at the FC geometry and the S$_1$ (A$_1$) state at the last point of the relaxed scan are given as insets.}
    \label{fig: NH3-BH3 B-N relaxed scan}
\end{figure}

To answer this question, we performed a relaxed scan along the \ce{B-N} bond length at the MP2 level of theory in the ground electronic state and used the obtained geometries to calculate the electronic energies for the different excited electronic states of interest with ADC(2). The resulting electronic energies for the first nine electronic states are represented in Fig.~\ref{fig: NH3-BH3 B-N relaxed scan}. 

At the FC point -- ground-state optimized geometry, indicated by the gray dashed line in Fig.~\ref{fig: NH3-BH3 B-N relaxed scan} -- the two degenerate lowest electronic states (E symmetry) have a $\sigma_{BH}\rightarrow$Ryd$_{3s}$ character. The $\sigma_{BH}$ is a linear combination of the $\sigma$ orbitals constituting the \ce{B-H} bonds, while Ryd$_{3s}$ is the Rydberg-like $3s$ virtual orbital on the nitrogen. The other higher excited states have a similar nature and are all characterized by the same donating orbital ($\sigma_{BH}$). However, the accepting orbital is either a Rydberg-like $3p$s orbital of the nitrogen (Ryd$_{3p}$) or a more complex orbital involving the $3s$ orbital on the boron (see Fig.~S1). The trend changes when looking at the ninth electronic state (A$_1$ symmetry), which, at the FC point, corresponds to a n$_{N}\rightarrow$Ryd$_{3s}$ transition, where the n$_{N}$ is the lone pair of the nitrogen. 

Now that electronic-state characters are defined at the FC point, let us follow the evolution of the electronic states along the \ce{B-N} coordinate. Upon elongation of this bond, we notice a strong destabilization of the $\sigma_{BH}\rightarrow$Ryd$_{3s}$ states, in stark contrast with the n$_{N}\rightarrow$Ryd$_{3s}$ state that becomes the first excited state for a \ce{B-N} bond longer than 2\AA. This change of electronic character between adiabatic electronic states is validated by monitoring the NTOs for the first excited state at the last point of the relaxed scan (see inset in Fig.~\ref{fig: NH3-BH3 B-N relaxed scan} at $\sim 3.6$\AA), which confirms the local excitation of ammonia from its lone pair to the Ryd$_{3s}$ orbital. This conclusion is further corroborated by noting that the energy gap is 6.26 eV between this excited electronic state and the ground state at the last point of the LIIC, a value consistent with the first excited state of \ce{NH3} -- calculated at 6.20 eV with the same level of electronic-structure theory.  

From a benchmark perspective, the D$_1$ diagnostic for the MP2 ground state remains well below the recommended limit value of 0.04\cite{janssen1998new} along the entire \ce{B-N} relaxed scan (dotted orange line in Fig.~\ref{fig: NH3-BH3 B-N relaxed scan}). This observation suggests that the ground state is still well described by a (closed-shell) single configuration, consistent with the suggested heterolytic dissociation. We also compared the electronic energies obtained with ADC(2) to those calculated with SA(13)-CASSCF(8/9) and XMS(13)-CASPT2(8/9) on the support of the relaxed scan (Fig.~S5). The three different methods offer similar trends for the behavior of the electronic states along the \ce{B-N} dissociation coordinate, and we highlight here the rather good agreement between ADC(2) and XMS-CASPT2 (despite XMS-CASPT2 not describing all the possible Rydberg states due to the limited active space employed). 

The presence of a n$_{N}\rightarrow$Ryd$_{3s}$ electronic state, leading to the weakening of the \ce{B-N} bond by removing an electron from the nitrogen lone pair, could offer an explanation for the photodissociation process observed in different experimental studies.\cite{matsuo2014photodissociation, matsumoto2015synthesis, matsumoto2017design, ando2021boron} 
If the ammonia borane molecule is photoexcited in its lowest E states (S$_1$ and S$_2$ in the FC region), an elongation of the \ce{B-N} could result in an adiabatic transfer to an n$_{N}\rightarrow$Ryd$_{3s}$ character occurring at $\sim$2\AA~(see Fig.~\ref{fig: NH3-BH3 B-N relaxed scan}), even if this process would require to overcome an activation barrier. Once the electronic character of the molecule evolved to that of the dissociative state, the \ce{H3N-BH3} should spontaneously dissociate, resulting in the formation of an electronically excited ammonia molecule and a borane in its ground state (see S$_1$ inset at long \ce{B-N} distance in Fig.~\ref{fig: NH3-BH3 B-N relaxed scan}), in agreement with the reaction channel (a). We note that pathway (e), leading to the formation of an excited borane and ground-state ammonia, is not observed in the range of electronic energies studied here. 

\subsubsection{Pathway (b): H photodissociation from the \ce{BH3} moiety}
\label{subsubsec: B-H diss}

\begin{figure}[h!]
    \centering
    \includegraphics[width=0.80\textwidth]{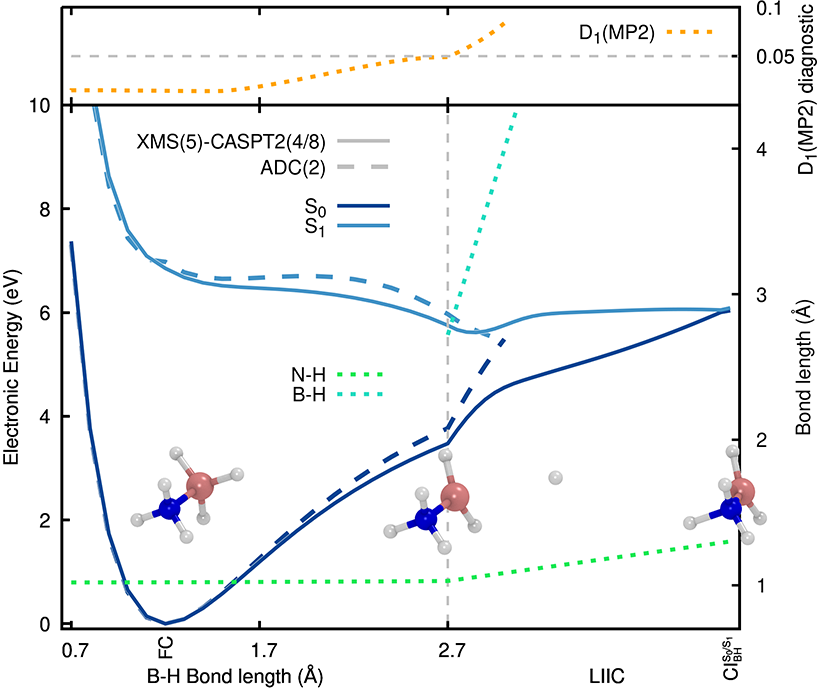}
    \caption{Electronic-energy profile along the H photodissociation of ammonia borane from the \ce{BH3} moiety. The profile is composed of a relaxed scan along the H dissociation coordinate (performed in the ground electronic state with MP2/aug-cc-pVDZ), combined with a LIIC that connects the last point of the relaxed scan to the geometry of the $\ciab{\text{BH}}$. The separation between the scan and the LIIC is indicated by a vertical dashed gray line. The two lowest electronic states are depicted in dark and light blue, respectively. ADC(2)/aug-cc-pVDZ and XMS(5)-CASPT2(4/8)/aug-cc-pVDZ profiles are highlighted by dashed and solid lines, respectively. The relevant B-H and the N-H bond lengths are shown in blue-green and green dotted lines along the profile. The D$_1$ diagnostic for the MP2/aug-cc-pVDZ ground state is reported as a dotted orange line in the upper panel. In the inset, we report three critical structures, namely the FC geometry, an intermediate geometry (the connection between the relaxed scan and the LIIC), and the $\ciab{\text{BH}}$ geometry. In the latter molecular representation, the leaving H atom is too far from the \ce{NH3-BH2} molecule to be visible.}
    \label{fig: NH3-BH3 B-H scan}
\end{figure}

We investigate here the H photodissociation occurring from the \ce{BH3} moiety, which we expect to be more favorable given the character of the first excited electronic state discussed above. We note that, in contrast with the \ce{B-N} stretch studied in the previous section, the \ce{B-H} and \ce{N-H} stretching modes are not totally symmetric and lift the degeneracy of the E states. As a result, we are not using labels with spatial symmetry in the following.

The \ce{B-H} bond cleavage is illustrated in Fig.~\ref{fig: NH3-BH3 B-H scan}, where a relaxed scan (for the ground state, MP2) is combined with a LIIC towards the optimized geometry for the \ce{B-H} dissociation ($\ciab{\text{BH}}$).
The first excited electronic state (S$_1$) is stabilized upon \ce{B-H} elongation, in agreement with its $\sigma_{BH}\rightarrow$Ryd$_{3s}$ character (see NTOs in Fig.~\ref{fig: NH3-BH3 B-N relaxed scan}), and a \ce{B-H} dissociation can proceed with almost no activation barrier, suggesting a fast photodissociation of the H atom from the \ce{BH3} moiety (Fig.~\ref{fig: NH3-BH3 B-H scan}). Up to the last point of the scan (indicated with a dashed vertical gray line in Fig.~\ref{fig: NH3-BH3 B-H scan}), the MP2/ADC(2) and XMS-CASPT2 electronic energies are in excellent agreement. At the last point of the scan, the \ce{B-H} length is 2.7\AA~(the sum of the van der Waals radii of H and B is $\sim$ 3.0\AA) while the S$_0$/S$_1$ energy gap is still $>$2 eV. This observation suggests that the \ce{B-H} bond is fully broken in the excited electronic state, and the \ce{B-H} elongation may not be the only coordinate that brings the molecule to the intersection region. 

How does the molecule reach the S$_1$/S$_0$ intersection after photodissociation of H from the \ce{BH3} moiety? The answer can be found by inspecting the LIIC connecting the relaxed scan to the $\ciab{\text{BH}}$ (Fig.~\ref{fig: NH3-BH3 B-H scan}, right side of the vertical dashed gray line). The $\ciab{\text{BH}}$ geometry is characterized by \ce{NH3-BH2} and an H atom completely dissociated (\ce{B-H} distance is $\sim$ 9\AA).  The \ce{NH3-BH2} molecule exhibits a longer \ce{N-H} bond (longer by $\sim$ 0.3\AA~with respect to the last point of the relaxed scan, see green dotted line in Fig.~\ref{fig: NH3-BH3 B-H scan}). This distortion of the \ce{NH3-BH2} moiety appears to take the molecule towards the $\ciab{\text{BH}}$ once the \ce{B-H} bond is cleaved in the first excited electronic state. The \ce{N-H} elongation process is somewhat captured by ADC(2), which shows a point of degeneracy between the S$_1$ and S$_0$ electronic energies. Still, care needs to be taken, as the S$_1$/S$_0$ crossing point occurs at a much shorter \ce{N-H} bond length when one compares the ADC(2) profile with that obtained with XMS-CASPT2. This fast interception of the S$_1$ state by S$_0$ is caused by an artificial destabilization of the MP2 ground state, clearly visible from the LIIC in Fig.~\ref{fig: NH3-BH3 B-H scan}. This conclusion is reinforced by the sharp increase of the D$_1$ diagnostic at the beginning of the LIIC, suggesting that the ground electronic state cannot be described adequately by a single closed-shell configuration. Hence, while the \ce{B-H} photodissociation of ammonia borane is well described by ADC(2), the subsequent geometrical distortions leading to the S$_1$/S$_0$ intersection region via the \ce{N-H} elongation push this method beyond its limits. 

It is worth noticing that the \ce{N-H} elongation discussed above does not necessarily imply that the \ce{N-H} bond will dissociate but may simply act as a distortion mode to access the intersection region. We will discuss later a different photodissociation channel involving an S$_1$/S$_0$ crossing mediated by an \ce{N-H} dissociation that occurs \textit{readily} after the \ce{B-H} dissociation. This process follows a similar route to what is depicted in Fig.~\ref{fig: NH3-BH3 B-H scan}, but would likely lead to the release of \ce{H2} and \ce{H2N=BH2} (as observed in the photodissociation of ethane\cite{chang2020ultraviolet}).

\subsubsection{Pathway (c): \ce{H2} dissociation from the \ce{BH3} moiety}
\label{subsubsec: H2 diss}

\begin{figure}[h!]
    \centering
    \includegraphics[width=0.75\textwidth]{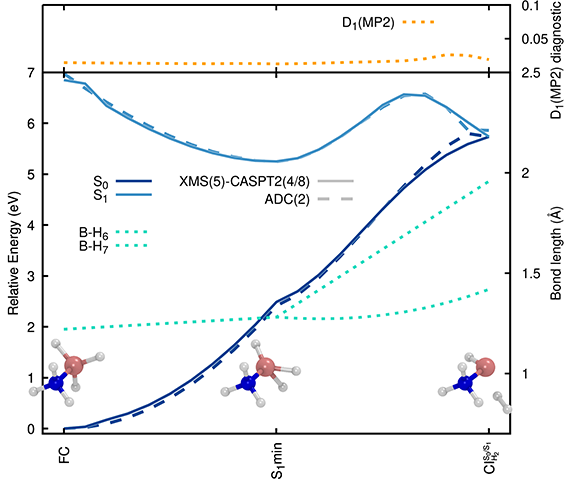}
    \caption{LIIC between the FC, S$_1$min, and $\ciab{\text{H}_\text{2}}$ geometries of ammonia borane. The two lowest electronic states are depicted in dark and light blue. ADC(2)/aug-cc-pVDZ and XMS(5)-CASPT2(4/8)/aug-cc-pVDZ electronic energies are reported in dashed and solid lines, respectively. The two relevant \ce{B-H} bonds are shown in blue-green dotted lines. The D$_1$ diagnostic for the MP2 ground state is reported as a dotted orange line in the upper panel. In the inset, we provide the three critical structures used to create the LIIC, namely the FC point, the S$_1$min, and the $\ciab{\text{H}_\text{2}}$.}
    \label{fig: NH3-BH3 H2 LIIC}
\end{figure}

The release of \ce{H2} from photoexcited ammonia borane can occur via the simultaneous dissociation of two H atoms from the \ce{BH3} moiety. To investigate this process, we perform a LIIC between the FC point optimized with MP2, the S$_1$min optimized with ADC(2), and the MECI for the \ce{H2} release ($\ciab{\text{H}_\text{2}}$). The electronic energies calculated with MP2/ADC(2) and XMS(5)-CASPT2(4/8) along the LIIC are shown in Fig.~\ref{fig: NH3-BH3 H2 LIIC}. 

The first part of the LIIC -- connecting the FC point to the S$_1$min -- is characterized by a Jahn–Teller distortion of the molecule along which two out of the three \ce{B-H} bonds elongate and distort. Such an elongation is visible from the \ce{B-H}$_6$ and \ce{B-H}$_7$ bond lengths displayed as dotted blue-green lines in Fig.~\ref{fig: NH3-BH3 H2 LIIC}, while the distortion is visible in the S$_1$min structure given as an inset. The S$_1$ electronic state is stabilized by $\sim$ 1.7 eV along this first segment of the LIIC. 

The second part of the LIIC -- connecting the S$_1$min to the $\ciab{\text{H}_\text{2}}$ -- highlights the concerted photodissociation of the two \ce{B-H} bonds and the release of \ce{H2}, leaving \ce{NH3-BH}. The $\ciab{\text{H}_\text{2}}$ geometry indicates that the \ce{H-H} bond is already formed when the S$_0$/S$_1$ electronic states cross. While one of the \ce{B-H} distances is still shorter than 1.5\AA~at the $\ciab{\text{H}_\text{2}}$ geometry, the interaction between the boron atom and the \ce{H2} molecule is presumably weak enough to allow the departure of the hydrogen molecule after a return to S$_0$. 
Upon inspection of the S$_1$ energy along the LIIC, we can expect that the photodissociation of ammonia borane into \ce{H2} and \ce{NH3-BH} is energetically feasible, given that the energy barrier along the LIIC is lower than the S$_1$ energy at the FC point. We note here that a LIIC does not provide a minimum-energy path and, as such, energy barriers along a LIIC are likely to be overestimated.

Comparing the ADC(2) and XMS-CASPT2 electronic energies along this LIIC reveals an excellent agreement between the two methods. This accuracy is also retained close to the CI region where the D$_1$ diagnostic remains surprisingly low. Hence, ADC(2) is expected to adequately describe the excited-state dynamics leading to the \ce{H2} release via the concerted breaking of two \ce{B-H} bonds. 

\subsubsection{Pathway (d): H dissociation from the \ce{NH3} moiety}
\label{subsubsec: N-H diss}
We finally note that electronic energies provided by ADC(2) agree well with those obtained with XMS-CASPT2 for the direct \ce{N-H} bond breaking upon photoexcitation. To evaluate the likelihood of this direct \ce{N-H} photodissociation, we performed a relaxed scan in the S$_1$ excited electronic state (with ADC(2)) along the \ce{N-H} bond of \ce{H3N-BH3} (see Fig.~S6). The energy gained during the relaxation following photoexcitation from the FC point to the S$_1$min (see Fig.~\ref{fig: NH3-BH3 H2 LIIC}) makes a direct \ce{N-H} photodissociation plausible. 

\subsubsection{Photoabsorption cross-section and nonadiabatic dynamics of ammonia borane}
\label{subsubsec: NH3BH3 photoabsorption cross-section and nonadiabatic dynamics}
The main summary of the previous section is all the proposed photodissociation channels could be accessible upon photoexcitation of ammonia borane, with the likelihood of reaching a given decay pathway being dictated by the initial excitation wavelength and subsequent coupled electron-nuclear dynamics between the excited electronic states. Hence, we now turn our attention to the determination of a photoabsorption cross-section for ammonia borane and the investigation of the potential photochemical pathways triggered at different excitation wavelengths using nonadiabatic molecular dynamics simulations. Based on the overall good agreement observed between ADC(2) and XMS-CASPT2 for the different photodissociation pathways (and keeping in mind the potential limitations of the former method, we decided to use the ADC(2)/aug-cc-pVDZ level of electronic structure theory for the photoabsorption cross-section of \ce{H3N-BH3} and its nonadiabatic molecular dynamics with TSH (see Sec.~\ref{sec: computational details} for additional details).  

\begin{figure}[h!]
    \centering
    \includegraphics[width=0.80\textwidth]{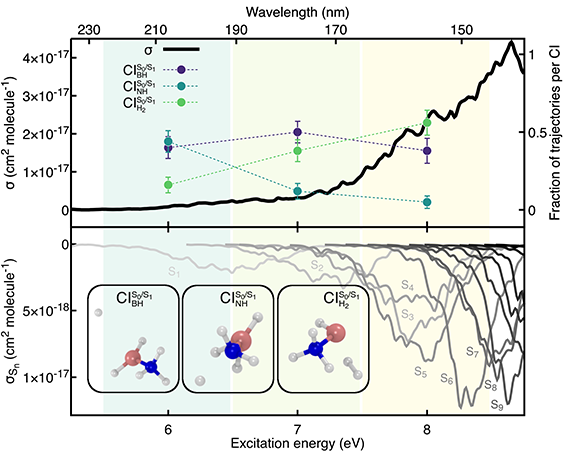}
    \caption{Photoabsorption cross-section and fractions of trajectories reaching a certain CI for \ce{H3N-BH3}. 
    Upper panel: full photoabsorption cross-section (black solid line). The spectral range is partitioned into three energy windows shown as green, light green, and yellow areas. The fraction of trajectories undergoing \ce{B-H} dissociation ($\ciab{\text{BH}}$, purple), \ce{N-H} dissociation ($\ciab{\text{NH}}$, blue-green) and \ce{H2} release from the cleavage of two \ce{B-H} bonds ($\ciab{\text{H}_\text{2}}$, light green) is determined for each energy window and reported with filled circles.
    Lower panel: individual excited-state contributions to the full photoabsorption cross-section depicted with solid lines colored from light gray (S$_0\rightarrow$S$_1$) to black (S$_0\rightarrow$S$_{15}$). Exemplary molecular structures for the three photodissociation pathways are shown as insets. ADC(2)/aug-cc-pVDZ is used for the electronic structure.}
    \label{fig: QY NH3BH3}
\end{figure}

The total photoabsorption cross-section of \ce{H3N-BH3} is shown in the upper panel of Fig.~\ref{fig: QY NH3BH3} (black solid line) and further decomposed into its different electronic-state contributions (S$_0\rightarrow$S$_n$) in the lower panel of the same figure. The first electronic transition (S$_0\rightarrow$S$_1$) spans a rather broad energy range (6-7 eV) and, together with a small contribution from S$_0\rightarrow$S$_2$ excitation, is the main transition present in the first selected energy window (shaded green area) for the subsequent excited-state dynamics simulations. The second and third energy windows (light-green and yellow areas) are more congested and contain electronic excitations to several different electronic states. The intensity of the photoabsorption cross-section absorption grows across the three windows, with S$_0\rightarrow$S$_6$ and S$_0\rightarrow$S$_9$ being the most intense bands in this energy range.

Having defined different regions of interest in the photoabsorption cross-section of ammonia borane, we can now proceed with the nonadiabatic dynamics by initiating a swarm of TSH trajectories in each of the three energy windows (green, light green, and yellow areas in Fig.~\ref{fig: QY NH3BH3}). Around 50 TSH trajectories were launched in each energy window (see Table S1 for the precise values). Our earlier analysis shows that the fraction of trajectories reaching a particular CI can be associated with a specific photodissociation pathway (upper panel of Fig.~\ref{fig: QY NH3BH3}). Three of the four expected channels discussed above were observed in the TSH dynamics: the \ce{B-H} dissociation ($\ciab{\text{BH}}$), \ce{N-H} elongation ($\ciab{\text{NH}}$), and the \ce{H2} release from the concerted cleavage of two \ce{B-H} bonds ($\ciab{\text{H}_\text{2}}$). Comparing the fraction of trajectories reaching each of these three CIs shows that the \ce{B-H} dissociation pathway appears to be largely independent of the excitation energy. In contrast, the concerted cleavage of two \ce{B-H} bonds becomes more favorable at higher excitation energies, at the expense of the nonradiative decay via the \ce{N-H} elongation. No direct \ce{B-N} bond photodissociation in the excited electronic states was observed in the TSH dynamics.

Since we showed that ADC(2) behaves properly when elongating the \ce{N-H} bond close to the CIs region (see Fig.~S6), we restarted five trajectories in S$_0$ after they reached the $\ciab{\text{NH}}$ and carried on 500 fs of adiabatic dynamics at the MP2/aug-cc-pVDZ level of theory. We observed a fast reformation of the \ce{N-H} bond in S$_0$ for all five trajectories, followed by an unexpected \ce{B-N} dissociation in S$_0$ due to a vibrationally hot \ce{H3N-BH3} adduct. This \ce{B-N} dissociation should not be confused with the photodissociation described in Sec.~\ref{subsubsec: NH3BH3 adduct photodissociation} where the \ce{B-N} bond breaking takes place in an excited electronic state and leads to the formation of an electronically excited \ce{NH3}.

Despite the surprisingly rich photochemistry of \ce{H3N-BH3}, the photodissociation of the adducts still remains elusive. Given the nature of the low-lying electronic states in \ce{H3N-BH3}, one may be tempted to play with substitutions on the \ce{NH3} or \ce{BH3} moiety to alter the ordering of the excited electronic states and potentially favor the \ce{B-N} bond rupture. 

\subsection{Photochemistry of pyridine borane}
\label{subsec: pyridine-borane}

Let us now investigate the photochemistry of pyridine borane, \ce{Py-BH3}. The overall motivations behind the modifications of the Lewis base (and later the Lewis base) forming the \ce{B-N} bond are to (i) lower the transition energies to make them more readily accessible for future spectroscopic studies and (ii) favor the excited photodissociation associated with the channel (a) from Fig.~\ref{fig: scheme photodissociation}, that is, the rupture in an excited electronic state of the \ce{B-N} bond. From a computational perspective, we will use MP2 and ADC(2) for all calculations, given their good performance for ammonia borane and the increase in computational cost resulting from the substitution of \ce{NH3} by pyridine.

The \ce{B-N} bond length of \ce{Py-BH3} is slightly shorter than that of \ce{H3N-BH3} at the optimized ground-state geometry: 1.64\AA~vs 1.66\AA, respectively (MP2/aug-cc-pVDZ, see Table~{S4}). This slight shortening of the \ce{B-N} bond, connected to the sp$^2$ hybridization of the nitrogen in pyridine, confirms that the interaction between the Lewis acid and base remains significant. This finding is further corroborated by gas-phase binding energies reported by Potter et al.\cite{potter2010thermochemistry}

\subsubsection{Photodissociation of the B--N bond in pyridine borane}
\label{subsubsec: PyBH3 adduct photodissociation}
\begin{figure}[h!]
    \centering
    \includegraphics[width=0.75\textwidth]{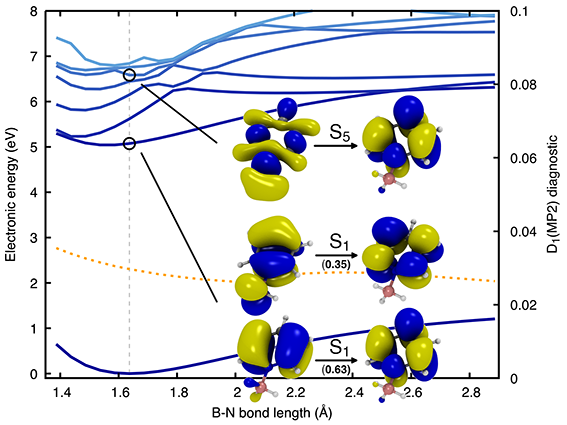}
    \caption{Relaxed scan along the \ce{B-N} bond length of \ce{Py-BH3} obtained for the ground electronic state with MP2. Excited electronic energies were obtained with ADC(2)/aug-cc-pVDZ. The FC point is indicated by the gray vertical dashed line. The lowest eight electronic states are calculated along the scan and depicted by lines in shades of blue -- from S$_0$ in dark blue to S$_7$ in light blue. The D$_1$ diagnostic for the MP2 ground state is reported as a dotted orange line. NTOs for the S$_1$ (two pairs of NTOs are required to describe the first excited state) and S$_5$ electronic state at the FC geometry are given as insets.}
    \label{fig: Py-BH3 B-N relaxed scan}
\end{figure}

As for \ce{H3N-BH3}, we performed a relaxed scan along the \ce{B-N} bond with MP2 and used these structures to calculate electronic energies with ADC(2) (see Fig.~\ref{fig: Py-BH3 B-N relaxed scan}). Focusing our attention on the FC geometry, we notice that the first excited state -- located at 5 eV above S$_0$ -- can be described by \textit{two} pairs of NTOs with the following character: $\pi\rightarrow\pi^{*}$ (lower NTO pair with a singular value equal to 0.63) with a small contribution of the type $\pi' + \sigma_{BH}\rightarrow\pi'^{*}$ (upper NTO pair with a singular value equal to 0.35). The $\sigma_{BH}$ contribution to the S$_1$ state is similar to the contribution observed for the first two excited states of \ce{H3N-BH3} (in the FC region). However, the first excited state of \ce{Py-BH3} is 2 eV lower than the first excited state of \ce{H3N-BH3} thanks to the $\pi$ system of \ce{Py}, meaning that Rydberg-like orbitals do not contribute to the low-lying electronic states of \ce{Py-BH3}. 

Looking at the NTOs describing the S$_5$ electronic state of \ce{Py-BH3} (inset in Fig.~\ref{fig: Py-BH3 B-N relaxed scan}), we notice that the donating orbital exhibits a clear signature of the pyridine lone pair (n$_{N}$). Similar to what was observed for \ce{H3N-BH3}, the character of this dissociative electronic state can be followed diabatically from S$_5$ in the FC region down to S$_1$ for longer \ce{B-N} distances. 

\subsubsection{Photoabsorption cross-section and nonadiabatic dynamics of pyridine borane}
\label{subsubsec: PyBH3 photoabsorption cross-section and nonadiabatic dynamics}

\begin{figure}[h!]
    \centering
    \includegraphics[width=0.80\textwidth]{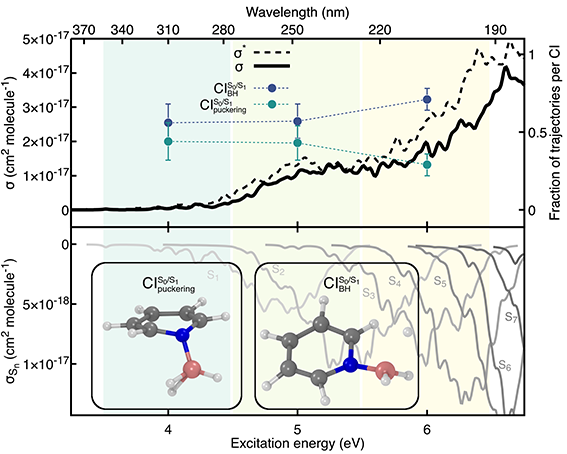}
    \caption{
    Photoabsorption cross-section and fractions of trajectories reaching a certain CI for \ce{Py-BH3}. 
    Upper panel: full photoabsorption cross-section obtained with ADC(2)/cc-pVDZ (solid black line) and ADC(2)/aug-cc-pVDZ (dashed black line). The spectral range is partitioned into three energy windows shown as green, light green, and yellow areas. The fraction of trajectories undergoing \ce{B-H} dissociation ($\ciab{\text{BH}}$, purple) or ring puckering dissociation ($\ciab{\text{puckering}}$, blue-green) is determined for each energy window and reported with filled circles.
    Lower panel: individual excited-state contributions to the full photoabsorption cross-section depicted with solid lines colored from light gray (S$_0\rightarrow$S$_1$) to dark gray (S$_0\rightarrow$S$_{8}$). Exemplary molecular structures for the two photodissociation pathways are shown as insets.}
    \label{fig: QY PyBH3}
\end{figure}

Mirroring Sec.~\ref{fig: QY NH3BH3} on ammonia borane, we now discuss the photoabsorption cross-section and nonadiabatic dynamics of pyridine borane. Before doing so, we start with a technical note on the basis set used for the upcoming calculations. Given the reduced presence of Rydberg states in the low-lying excited states of this Lewis adduct, we used the smaller basis set cc-pVDZ instead of aug-cc-pVDZ. Vertical transition energies obtained with ADC(2)/cc-pVDZ are in good agreement with those calculated with ADC(2)/aug-cc-pVDZ (see Table S3 in the ESI). To further confirm the validity of removing the diffuse functions, we calculated the photoabsorption cross-section using ADC(2)/cc-pVDZ and ADC(2)/aug-cc-pVDZ (solid and dashed black lines in the top panel of Fig.~\ref{fig: QY PyBH3}). Once again, the two cross-sections show a good overlap, particularly for excitation energies below 6.5 eV. The ADC(2)/aug-cc-pVDZ appears to be only slightly red-shifted with respect to the ADC(2)/cc-pVDZ spectra.  

The first energy window (green area in Fig.~\ref{fig: QY PyBH3}) contains only the weak low-energy tail coming from the first excited state with excitation energies extending up to 4.5 eV. The second and third windows (light-green and yellow areas in Fig.~\ref{fig: QY PyBH3}, respectively) cover more electronic transitions -- the second window, in particular, contains transitions towards S$_1$, S$_2$ and S$_3$. 

The nonadiabatic dynamics simulation of \ce{Py-BH3} reveals two dominant nonradiative pathways. The first one involves the cleavage of a \ce{B-H} bond, consistent with the character of the electronic state discussed in Sec.~\ref{subsubsec: PyBH3 adduct photodissociation} and in line with the $\ciab{\text{BH}}$ observed for \ce{H3N-BH3}. The second nonradiative pathway, characterized by the $\ciab{\text{puckering}}$ in Fig.~\ref{fig: QY PyBH3}, corresponds to the puckering of the pyridine ring. Interestingly, the puckering motion either involves the nitrogen atom (shown in the inset of Fig.~\ref{fig: QY PyBH3}) -- provoking an out-of-plane motion of the \ce{BH3} moiety -- or the distortion of one or more carbon atoms of the ring. This nonradiative process closely resembles the nonradiative relaxation mechanism of isolated pyridine, discussed in detail in the literature.\cite{varras2018explanation, varras2020can} The concerted dissociation of two \ce{B-H} bonds is not observed in the excited-state dynamics conducted for \ce{Py-BH3}. However, we note that we cannot rule out a release of \ce{H2} upon relaxation of \ce{Py-BH3} to the ground state or following photoexcitation at higher energies. 

Analysis of the wavelength dependence of the two deactivation channels reveals that the fraction of trajectories suffering a \ce{B-H} dissociation or puckering is about the same for the first and second energy windows. However, the fraction of trajectories reaching the $\ciab{\text{BH}}$ increases substantially at higher excitation energies at the expense of the puckering relaxation process. 

The substitution of \ce{NH3} with \ce{Py} has dramatically changed the electronic structure and the underlying photochemistry of the \ce{B-N} Lewis adducts. However, a \ce{B-N} photodissociation pathway, yet again, remains elusive for this molecule at the excitation wavelength that we probed. Therefore, we propose an additional alteration of our model Lewis adduct by tuning its \ce{B-N} bond strength. Weakening the strength of the \ce{B-N} bond should, in principle, imply that the excited state with an n$_{N}$ character decreases in energy. To achieve this goal, we used a weaker Lewis acid, boric acid \ce{B(OH)3}, instead of the stronger \ce{BH3}.

\subsection{Photochemistry of pyridine-boric acid}
\label{subsec: pyridine-boric acid}
The last \ce{B-N} model studied here is pyridine--boric acid, \ce{Py-B(OH)3}. \ce{Py-B(OH)3} exhibits a longer \ce{B-N} bond at its ground-state optimized geometry (1.73\AA) than that of \ce{H3N-BH3} (1.66\AA) and \ce{Py-BH3} (1.64\AA), in line with the expected weaker \ce{B-N} bond for this molecule. To explore whether the weaker nature of the \ce{B-N} bond in the ground electronic state has implications for the excited electronic properties of the molecule, we performed the same type of calculations as done for the other \ce{B-N} adducts, starting by monitoring the electronic energies along a \ce{B-N} relaxed scan. 

\subsubsection{Photodissociation of the B--N bond in pyridine-boric acid}
\label{subsubsec: PyBoricacid adduct photodissociation}
\begin{figure}[h!]
    \centering
    \includegraphics[width=0.75\textwidth]{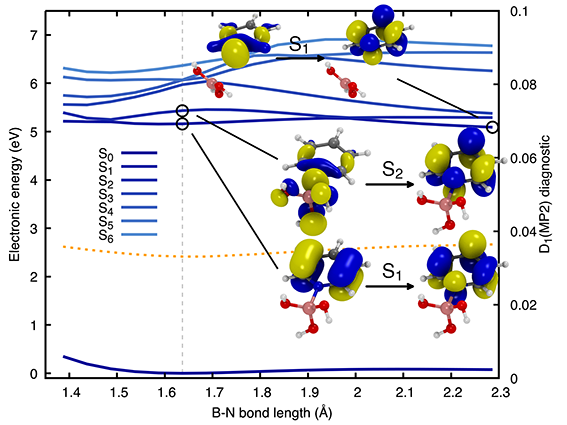}
    \caption{
    Relaxed scan along the \ce{B-N} bond length of \ce{Py-B(OH)3} obtained for the ground electronic state with MP2/aug-cc-pVDZ. Excited electronic energies were obtained with ADC(2)/aug-cc-pVDZ. The FC point is indicated by the gray vertical dashed line. The lowest eight electronic states are calculated along the scan and depicted by lines in shades of blue -- from S$_0$ in dark blue to S$_7$ in light blue. The D$_1$ diagnostic for the MP2 ground state is reported as a dotted orange line. NTOs for the S$_1$ and S$_2$ excitations at the FC geometry are given as insets.}
    \label{fig: Py-Boric Acid B-N relaxed scan}
\end{figure}

The relaxed scan obtained at the MP2 level of theory along the \ce{B-N} bond of \ce{Py-B(OH)3} is shown in Fig.~\ref{fig: Py-Boric Acid B-N relaxed scan}. The variation of the ground-state electronic energy along the \ce{B-N} relaxed scan highlights the weak nature of the \ce{B-N} bond. The S$_0$ electronic energy curve displays only a shallow minimum along this coordinate, in stark contrast with the minima observed earlier for \ce{H3N-BH3} and \ce{Py-BH3}. 

Focusing now on the excited electronic states at the FC geometry, we observe that the S$_2$ electronic state exhibits an n$_{N}\rightarrow\pi^*$ state (see NTOs, inset of Fig.~\ref{fig: Py-Boric Acid B-N relaxed scan}), again in line with a weaker \ce{B-N} bond for this adduct. The S$_2$ character at the FC geometry is similar to the n$_{N}\rightarrow\pi^*$ states of \ce{Py-BH3}, with an additional contribution coming from the lone pairs of the oxygen atoms on the boric acid moiety and a weaker contribution from the $\sigma$ \ce{C-C} and \ce{C-H} bonds of the pyridine ring. 

Following the n$_{N}\rightarrow\pi^*$ character along the \ce{B-N} relaxed scan confirms its stabilization upon a \ce{B-N} bond elongation, in line with our earlier observation for \ce{Py-BH3}. The diabatic crossing between the n$_{N}\rightarrow\pi^*$ and the $\pi\rightarrow\pi^*$ states occurs fairly close to the FC point ($\sim$1.9\AA), in contrast with \ce{Py-BH3}. This observation may imply that the n$_{N}\rightarrow\pi^*$ state can be populated more easily in \ce{Py-B(OH)3} than in \ce{Py-BH3} during the nonadiabatic dynamics. 

\subsubsection{Photoabsorption cross-section and nonadiabatic dynamics of pyridine-boric acid}
\begin{figure}[h!]
    \centering 
    \includegraphics[width=0.80\textwidth]{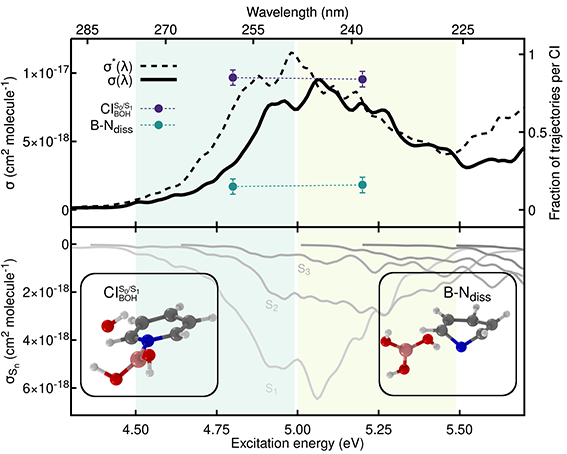}
    \caption{
    Photoabsorption cross-section and fractions of trajectories reaching a certain CI for \ce{Py-B(OH)3}. 
    Upper panel: full photoabsorption cross-section obtained with ADC(2)/cc-pVDZ (solid black line) and ADC(2)/aug-cc-pVDZ (dashed black line). The spectral range is partitioned into two energy windows shown as green and light green areas. The fraction of trajectories showing a B-OH dissociation ($\ciab{\text{BOH}}$, purple) or a \ce{B-N} dissociation (\ce{B-N}$_{diss}$, green) is determined for each energy window and reported with filled circles.
    Lower panel: individual excited-state contributions to the full photoabsorption cross-section depicted with solid lines colored from light gray (S$_0\rightarrow$S$_1$) to dark gray (S$_0\rightarrow$S$_{8}$). Exemplary molecular structures for the two photodissociation pathways are shown as insets.}
    \label{fig: QY PyBoricadic}
\end{figure}

We calculated the photoabsorption cross-section of \ce{Py-B(OH)3} and split it into two energy windows for the subsequent TSH nonadiabatic dynamics simulations (Fig.~\ref{fig: QY PyBoricadic}). The first window aims to investigate the photochemistry of the molecules triggered by a photoexcitation in the S$_1$ electronic state, with only minor contributions from excitation to the second excited electronic state S$_2$. The higher excitation window is instead more complex as it incorporates multiple contributions from broad electronic-transition bands (see lower panel of Fig.~\ref{fig: QY PyBoricadic}). These broad energy range for the electronic transitions are linked to the weak \ce{B-N} bond and, consequently, to the spread of the \ce{B-N} distance in the sampled geometries used to calculate the photoabsorption cross-section (see Fig.~S3).

The nonadiabatic dynamics simulations reveal for both energy windows a nonradiative pathway involving a \ce{B-OH} cleavage (purple circles in Fig.~\ref{fig: QY PyBoricadic}) -- analogous to the \ce{B-H} dissociation observed in the nonadiabatic dynamics of \ce{Py-BH3}. In addition, photoexcitation of \ce{Py-B(OH)3} resulted in direct \ce{B-N} photodissociation (green circles in Fig.~\ref{fig: QY PyBoricadic}). The \ce{B-N} dissociation occurs in a few tens of femtoseconds and the \ce{B(OH)3} fragment is released in its ground electronic state. In contrast, the pyridine moiety is produced in an excited electronic state -- in line with the results obtained in the relaxed scan presented above (Fig.~\ref{fig: Py-Boric Acid B-N relaxed scan}). The subsequent excited-state dynamics observed for pyridine is in line with earlier studies, where the molecule relaxes towards the ground state \textit{via} a puckering motion. The puckering of pyridine occurring after the \ce{Py-B(OH)3} photodissociation should not be confused with the puckering mechanism observed in the excited-state dynamics of \ce{Py-BH3}, as the \ce{B-N} bond in the latter was still intact (see Fig.~\ref{fig: QY PyBH3}). We note that the fast nonradiative decay of pyridine after the \ce{B-N} photodissociation is expected to suppress any sizable fluorescence signal. 
Photodissociable Lewis adducts, decaying to the ground state via nonradiative mechanisms, have already been discussed in the literature in line with our findings.\cite{kano2005photoswitching} It is worth noting that the photodissociation of \ce{Py-B(OH)3} results in an excited Lewis base, but depending on the electronic states available for the Lewis base or acid, one may also expect that the photodissociation would result to an excited Lewis acid (pathway (e) in Fig.~\ref{fig: scheme photodissociation}). Such a process was actually observed experimentally for conjugated Lewis acids, and fluorescence of both the adduct and the isolated excited Lewis acid were measured. \cite{matsuo2014photodissociation, matsumoto2015synthesis, matsumoto2017design, ando2021boron}   

\section{Implications of the theoretical findings for the interpretation of earlier photochemical experiments on \ce{B-N} Lewis adducts}
\label{sec: implications of the Lewis adduct photodissociation}
In the following, we propose to confront our newly gained understanding of the photophysics/photochemistry of \ce{B-N} Lewis adducts to earlier experimental works. In particular, we will be looking to extract potential guiding principles for predicting the photochemistry of a given \ce{B-N} Lewis adduct. 

\subsection{B--R photodissociation}
An example of the photolysis of a Lewis adduct involving the rupture of the bond between the boron and its substituent is given by the photochemistry of tribenzylborane-ammonia adducts.\cite{chung1976photochemistry} Irradiating the tribenzylborane-ammonia complex in the UV led to an efficient and heterolytic photodissociation of the benzyl carbon-boron bond. This process mirrors the \ce{B-H} photodissociation (pathway (b)) observed for \ce{H3N-BH3} and \ce{Py-BH3}. The authors of this study also highlighted the formation of bibenzyl upon irradiation of the uncoordinated tribenzylborane, while only traces of this photoproduct were observed in the photochemistry of the tribenzylborane-ammonia adduct. The formation of bibenzyl could possibly be connected to the pathway (c) associated with the formation of \ce{H2} in the photochemistry of \ce{H3N-BH3}, despite its small predicted quantum yield. 
Based on this comparison and our theoretical results, we may infer that a boron-based Lewis acid not embedded in a molecular framework might be prone to \ce{B-R} bond breaking upon light irradiation.  

\subsection{B--N photodissociation}
Shi et al. synthesized a series of molecules exhibiting frustrated, dynamic, and allowed \ce{B-N} pairs. Among these molecules, two allowed \ce{B-N} coordinated complexes display dative bond photodissociation.\cite{shi2022dynamic} The \ce{B-N} bond lengths of two of these complexes, one reversible and one allowed, are 1.805(2)\AA~and 1.792(3)\AA, respectively. These bond lengths are comparable with the \ce{B-N} distance in our \ce{Py-B(OH)3} adduct. Both of these complexes show photodissociation. However, the reversible complex also exhibits a thermodynamic cleavage of the \ce{B-N} bond in the ground electronic state at modest temperatures (T $> 298$ K), while the allowed complex stays bounded for all temperatures under investigation (T < $371$ K). These observations could be connected to our computational finding on \ce{Py-B(OH)3}, which shows \ce{B-N} photodissociation but is also prone to dissociate thermally due to the weak nature of its \ce{B-N} bond. Furthermore, both the reversible and allowed Lewis adducts display only a weak emission in THF, due to the presence of efficient nonradiative pathways. Yet, highly enhanced emission properties were observed in a THF/\ce{H2O} solution due to aggregation-induced emission effect and suppression of puckering and motions. This observation aligns with our findings that, upon \ce{B-N} dissociation, one of the moieties forming the original Lewis adduct remains in an excited electronic state, and the availability of nonradiative decay pathways will dictate its emissive behavior (in the gas phase). 

Matsuo et al., investigating the photochemistry of a planarized trinaphthylborane coordinated to pyridine, observed an unexpected dual fluorescence signal.\cite{matsuo2014photodissociation} The dual fluorescence was rationalized by the emission from (\textit{i}) the excited trinaphthylborane\ce{-Py} adduct and (\textit{ii}) the excited trinaphthylborane formed following the excited-state photodissociation of the Lewis adduct. The latter process appears to follow the pathway (e) highlighted in Fig.~\ref{fig: scheme photodissociation} and contrasts with the sole pathway (a) observed for ammonia borane and \ce{Py-B(OH)3}, where the Lewis base remained excited. The switch to pathway (e) for this molecule can be understood by the highly conjugated nature of the Lewis acid in trinaphthylborane. The findings by Matsuo and coworkers also suggest that, upon photoexcitation, the excited-state dissociation of the adduct was in competition with its decay via fluorescence (with a lifetime estimated to be 2.3 ns). 
One may therefore expect a rather slow excited-state dissociation process for the trinaphthylborane\ce{-Py}, in comparison to the photodissociation observed in \ce{Py-B(OH)3}, which takes place in a few hundreds of femtoseconds. This observation could be rationalized by two different effects. First, the trinaphthylborane\ce{-Py} adduct appears to have a slightly stronger \ce{B-N} bond ($\sim$ 1.69\AA~based on the X-ray structure obtained in Ref.~\citenum{matsuo2014photodissociation}, in line with the theoretical value of 1.69\AA~obtained with MP2/cc-pVDZ) than \ce{Py-B(OH)3} ($\sim$ 1.73\AA), which could make the excited-state photodissociation process less favorable. In addition, the photodissociation of the trinaphthylborane\ce{-Py} adduct completely disappears at a lower temperature (T $\sim$ 193 K). Based on our theoretical analysis, the required \ce{B-N} stretching necessary for the diabatic transition to occur and trigger the adduct photodissociation may not be possible at low kinetic energy or for too strong \ce{B-N} bonds. Second, the trinaphthylborane\ce{-Py} adduct does not possess efficient nonradiative decays, in contrast with \ce{H3N-BH3} and \ce{Py-BH3}. This property allows the adduct to remain in the excited electronic state for long enough to either slowly photodissociate or decay via fluorescence. This slow process is not possible in \ce{H3N-BH3} and \ce{Py-BH3} due to the presence of the other deactivation channels (pathways (b)-(d)).

\subsection{Tuning the B--N bond}
Voegtle et al. investigated the strength of Lewis adducts created by combining \ce{BF3} to a series of quinoline derivatives, acting as photoactive Lewis bases.\cite{voegtle2022can} The focus of this work was not primarily on the photochemistry of the Lewis adducts but on how different quinoline-based Lewis bases could affect the strength of the \ce{B-N} bond. A first observation from this work is that all quinolines showed an increased bonding affinity for \ce{BF3} when the adduct was in its first excited electronic state. This observation is consistent with the computational results obtained for our three models of Lewis adducts (in particular ammonia borane, see Fig.~\ref{fig: NH3-BH3 B-N relaxed scan}, for which the first excited electronic state is stabilized upon contraction of the \ce{B-N} bond). The authors also highlighted a correlation between the Hammett parameter of the substituent groups and the ability of quinoline to coordinate more strongly to \ce{BF3} upon photoexcitation. This observation was linked to the change in electron density on the nitrogen atoms, being either depleted or enhanced by the functionalization with electron-withdrawing or electron-donating groups. A quinoline functionalized with an electron-donating group was a stronger Lewis photobase (i.e., binding more strongly \ce{BF3}) than a quinoline functionalized with an electron-withdrawing group. This correlation resonates with our results, as an electron-rich acid (such as \ce{B(OH)3}) acts as a weaker photoacid (i.e., binding less strongly a pyridine) than an electron-deficient acid (such as \ce{BH3}). This can be qualitatively appreciated by comparing Figs.~\ref{fig: Py-BH3 B-N relaxed scan} and \ref{fig: Py-Boric Acid B-N relaxed scan}. In the former, the S$_1$ state is tangibly stabilized upon contraction of the \ce{B-N} bond (i.e., \ce{BH3} acts as moderate Lewis photoacid). In contrast, the S$_1$ state remains shallow for the \ce{Py-B(OH)3} adduct, and only a minute stabilization is observed (i.e., \ce{B(OH)3} acts as a weak Lewis photoacid). We finally note that some \ce{B-N} Lewis adducts may exhibit electronic states in which the \ce{B-N} bond is weakened due to the decrease (increase) of the electron density on the nitrogen (boron).

\section{Conclusion}
\label{sec: conclusion}
In this work, we studied three different \ce{B-N} Lewis adducts -- \ce{H3N-BH3}, \ce{Py-BH3}, \ce{Py-B(OH)3} -- using a broad range of techniques in computational photochemistry. We used ammonia borane, \ce{H3N-BH3}, as a model system to decipher the different nonradiative pathways of this class of molecules. This simple Lewis adduct exhibits a plethora of possible photochemical processes that can be accessed by using different excitation wavelengths. In particular, we observed a \ce{B-H} bond cleavage and a concerted bond breaking of two \ce{B-H} bonds, combined with nonradiative decays triggered by \ce{N-H} bond stretching. A similar \ce{B-H} bond cleavage in the excited state was also observed for the \ce{Py-BH3} adduct, together with puckering distortions characteristic of the photochemistry of pyridine. Interestingly, the excited-state dynamics simulations for \ce{H3N-BH3} and \ce{Py-BH3} did not lead to a direct photodissociation of the \ce{B-N} bond for the excitation wavelengths simulated. The adduct photodissociation was only observed for \ce{Py-B(OH)3}, which presents a very weak \ce{B-N} bond. The \ce{B-N} photodissociation in \ce{Py-B(OH)3} competes with a phototriggered \ce{B-OH} cleavage, in analogy with the \ce{B-H} photodissociation mechanism of \ce{H3N-BH3} and \ce{Py-BH3}.   

Overall, our simulations describe many possible photochemical pathways for \ce{B-N} Lewis adducts. In addition to the excitation wavelength, the importance of each pathway for a given \ce{B-N} Lewis adduct will be dictated by the precise molecular nature of the Lewis acid and base and the strength of the adduct, as exemplified in our rationalization of the experimental results on these molecular systems available in the literature.  If the Lewis acid forming the adduct possesses weak \ce{B-R} bonds, they are likely to be readily photolyzed. In the absence of (\textit{i}) a possible \ce{B-R} bond dissociation and (\textit{ii}) a nonradiative decay from the Lewis base or acid, \ce{B-N} photodissociation could take place, further enhanced if the \ce{B-N} bond is weak or there is sufficient internal energy to reach a dissociative character diabatically. The \ce{B-N} photodissociation will lead to a photoexcited Lewis acid or Lewis base, depending on the stabilization of the excited electronic states of each moiety. 

We believe that the possible photochemical routes highlighted in this work for \ce{B-N} Lewis adducts will provide insights for the interpretation of upcoming photochemical experiments on this family of molecules and hopefully stimulate future photodynamics studies employing ultrafast spectroscopic techniques.

\begin{acknowledgement}
The authors thank Prof Mike N. R. Ashfold for stimulating discussions on this work. This project has received funding from the European Research Council (ERC) under the European Union's Horizon 2020 research and innovation program (Grant agreement No. 803718, project SINDAM) and the EPSRC Grant EP/V026690/1.  
\end{acknowledgement}

\begin{suppinfo}

The Supporting Information contains a detailed electronic-structure benchmark, raw results for the excited-state trajectories, active-space orbitals, distribution of \ce{B-N} bond lengths for the adducts studied in this work. (PDF) The xyz geometries for the LIICs and scans presented in this work are provided. (ZIP)

\end{suppinfo}

\providecommand{\latin}[1]{#1}
\makeatletter
\providecommand{\doi}
  {\begingroup\let\do\@makeother\dospecials
  \catcode`\{=1 \catcode`\}=2 \doi@aux}
\providecommand{\doi@aux}[1]{\endgroup\texttt{#1}}
\makeatother
\providecommand*\mcitethebibliography{\thebibliography}
\csname @ifundefined\endcsname{endmcitethebibliography}
  {\let\endmcitethebibliography\endthebibliography}{}

\end{document}


\section{Supplementary Tables}

\begin{table}[ht]
\caption{Summary of the outcomes of the TSH simulations for the three Lewis adducts investigated in this work.}
\label{tab: tsh_summary}
\begin{tabular}{@{}llll}
\toprule
                            & \multicolumn{3}{c}{\ce{H3N-BH3}}  \\
\midrule
window:                     & 6\,eV     & 7\,eV     & 8\,eV     \\ 
\midrule
$\ciab{\text{BH}}$          & 13        & 17        & 13        \\
$\ciab{\text{H}_\text{2}}$  & 7         & 18        & 22        \\
$\ciab{\text{NH}}$          & 25        & 13        & 4         \\
\ce{N-B}$_{diss}$           & 0         & 0         & 0         \\
discarded                   & 5         & 8         & 13        \\
total                       & 50        & 56        & 52        \\
\toprule
                            & \multicolumn{3}{c}{\ce{Py-BH3}}   \\
\midrule
window:                     & 4\,eV     & 5\,eV     & 6\,eV     \\ 
\midrule
$\ciab{\text{BH}}$          & 10        & 12        & 29        \\
$\ciab{\text{puckering}}$   & 8         & 9         & 12        \\
\ce{N-B}$_{diss}$           & 0         & 0         & 0         \\
discarded                   & 9         & 9         & 21        \\
total                       & 27        & 30        & 52        \\
\toprule
                            & \multicolumn{3}{c}{\ce{Py-B(OH)3}}\\
\midrule
window:                     & 4.75\,eV  & 5.25\,eV  &           \\ 
\midrule
$\ciab{\text{BOH}}$         & 50        & 38        &           \\
\ce{N-B}$_{diss}$           & 9         & 7         &           \\
discarded                   & 5         & 15        &           \\
total                       & 64        & 60        &           \\
\bottomrule
\end{tabular}
\end{table}

\begin{table}[ht]
\centering
\caption{Excitation energies in eV (rows with gray background) and oscillator strengths (rows with white background) of ammonia borane computed at the ground-state optimized geometry (SCS-MP2/def2-SVPD). The excited electronic states are labeled according to the C$_{\text{3v}}$ point group.}
\begin{adjustbox}{width=1\textwidth}
\begin{tabular}{ c|c c c c c c} 
    \toprule
                        & S$_1$/S$_2$ (E)   & S$_3$/S$_4$ (E)   & S$_5$ (A$_2$) & S$_6$ (A$_1$)     & S$_7$/S$_8$ (E)   & S$_9$ (A$_1$)\\
    \hline
    ADC(2)              &               &               &               &                   &                   &           \\
    \rowcolor{gray!25}
    aug-cc-pVDZ         & 6.96          & 8.24          & 8.27          & 8.33              & 8.45              & 8.65      \\
                        & (0.013)       & (0.029)       & (0.0)         & (0.005)           & (0.069)           & (0.126)   \\
    \rowcolor{gray!25}
    aug-cc-pVTZ         & 7.07          & 8.34          & 8.39          & 8.42              & 8.55              & 8.82      \\
                        & (0.014)       & (0.036)       & (0.0)         & (0.005)           & (0.062)           & (0.129)   \\
    \hline
    SCS-ADC(2)          &               &               &               &                   &                   &           \\
    \rowcolor{gray!25}
    aug-cc-pVDZ         & 7.18          & 8.40          & 8.45          & 8.53              & 8.63              & 8.86      \\
                        & (0.017)       & (0.043)       & (0.0)         & (0.004)           & (0.055)           & (0.135)   \\
    \rowcolor{gray!25}
    aug-cc-pVTZ         & 7.31          & 8.52          & 8.59          & 8.64              & 8.75              & 9.04      \\
                        & (0.017)       & (0.051)       & (0.0)         & (0.004)           & (0.048)           & (0.138)   \\
    \hline
    EOM-CCSD            &               &               &               &                   &                   &           \\
    \rowcolor{gray!25}
    aug-cc-pVTZ         & 6.99          & 8.26          & 8.31          & 8.33              & 8.49              & 8.87      \\
                        & (0.014)       & (0.033)       & (0.0)         & (0.005)           & (0.063)           & (0.132)   \\
    \rowcolor{gray!25}
    aug-cc-pVQZ         & 7.02          & 8.31          & 8.35          & 8.39              & 8.51              & 8.91      \\
                        & (0.014)       & (0.033)       & (0.0)         & (0.005)           & (0.060)           & (0.130)   \\
    \hline
    XMS(13)-CASPT2(8/9) &               &               &               &                   &                   &           \\
    \rowcolor{gray!25}
    aug-cc-pVDZ         & 6.81          & 8.14          & 8.16          & 8.20              & 8.40              & 8.71      \\
                        & (0.010)       & (0.004)       & (0.0)         & (0.0)             & (0.060)           & (0.140)   \\
    \bottomrule
\end{tabular}
\end{adjustbox}
\label{tab: NH3-BH3 excitation energies}
\end{table}

\begin{table}[ht]
\centering
\caption{Excitation energies in eV (rows with gray background) and oscillator strengths (rows with white background) of pyridine borane computed at the ground-state optimized geometry (MP2/aug-cc-pVDZ). The excited electronic states are labeled according to the C$_{\text{s}}$ point group.}
\begin{tabular}{ c|c c c c c c} 
    \toprule
                        & S$_1$ (A'')   & S$_2$ (A'')   & S$_3$ (A')    & S$_4$ (A'')       & S$_5$ (A') \\
    \hline
    ADC(2)              &               &               &               &                   &           \\
    \rowcolor{gray!25}
    cc-pVDZ             & 5.16          & 5.82          & 6.34          & 6.71              & 6.75      \\
                        & (0.026)       & (0.150)       & (0.003)       & (0.198)           & (0.003)   \\
    \rowcolor{gray!25}
    aug-cc-pVDZ         & 5.07          & 5.62          & 6.15          & 6.44              & 6.58      \\
                        & (0.031)       & (0.156)       & (0.003)       & (0.202)           & (0.004)   \\
    \rowcolor{gray!25}
    aug-cc-pVTZ         & 5.02          & 5.53          & 6.04          & 6.38              & 6.53      \\
                        & (0.032)       & (0.142)       & (0.003)       & (0.193)           & (0.004)   \\
    \bottomrule
\end{tabular}
\label{tab: Py-BH3 excitation energies}
\end{table}

\begin{table}
\centering
\caption{\ce{B-N} bond lengths (\AA) calculated at the optimized ground-state geometry for \ce{H3N-BH3}, \ce{Py-BH3}, and \ce{Py-B(OH)3} Lewis adducts (MP2/aug-cc-pVDZ).} 
\begin{tabular}{ |c|c c c |} 
    \hline
                        & \ce{H3N-BH3}  & \ce{Py-BH3}   &\ce{Py-B(OH)3}     \\
    \hline
    B-N                 & 1.66          & 1.64          & 1.73              \\
    \hline
\end{tabular}
\label{tab: geometrical parameters at the FC point}
\end{table}

\section{Supplementary Figures}
\label{compdet}

\begin{figure}
    \centering
    \includegraphics[width=\linewidth]{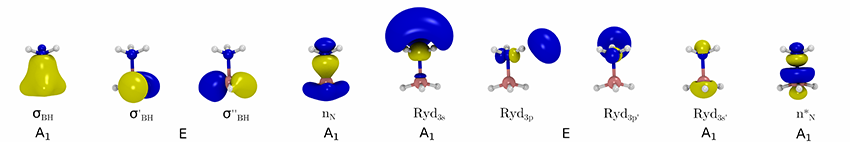}
    \caption{Representation of the SA(13)-CASSCF(8/9) natural orbitals and their labeling.}
\end{figure}

\begin{figure}
    \centering
    \includegraphics[width=\linewidth]{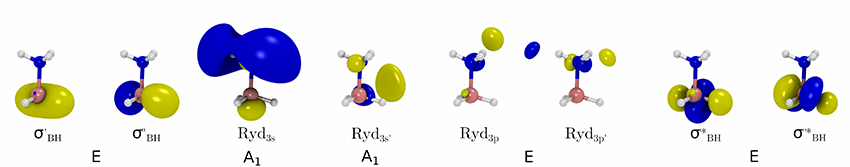}
    \caption{Representation of the SA(5)-CASSCF(4/8) natural orbitals and their labeling.}
\end{figure}

\begin{figure}
    \centering
    \includegraphics[width=\linewidth]{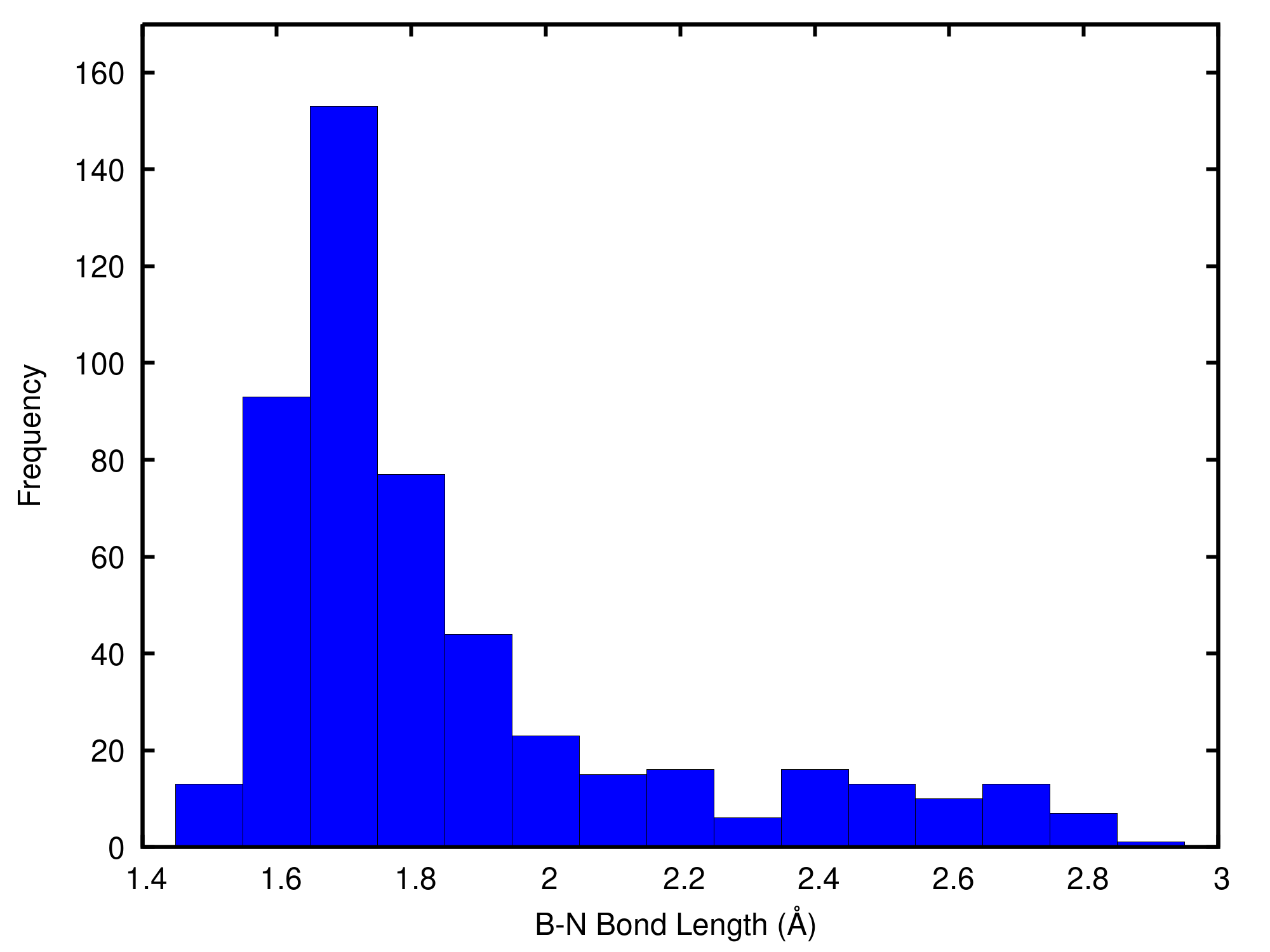}
    \caption{Distribution of the \ce{B-N} distances for \ce{Py-B(OH)3} collected during the QT ab initio molecular dynamics in the ground electronic state.}
\end{figure}

\begin{figure}
    \centering
    \includegraphics[width=\linewidth]{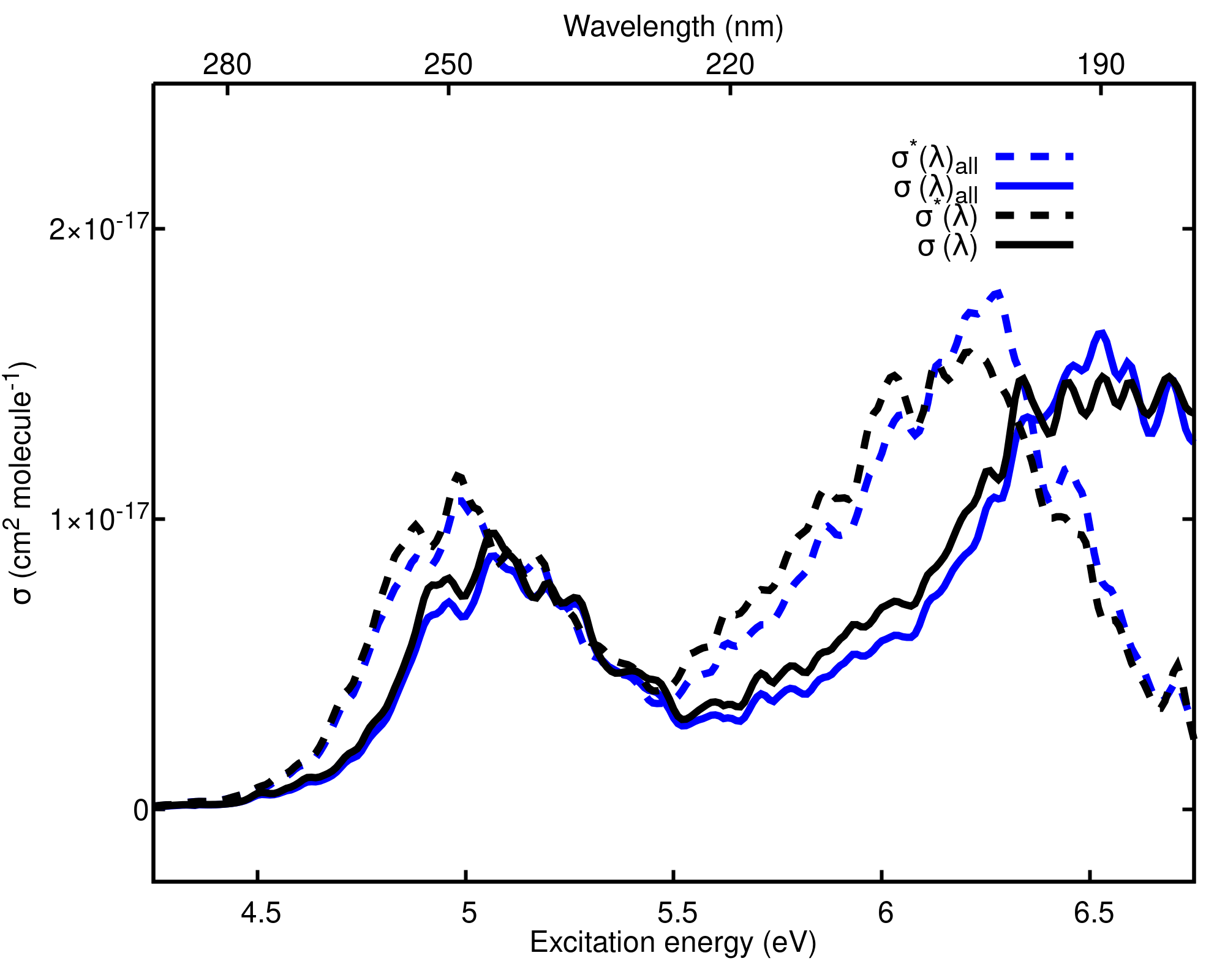}
    \caption{Photoabsorption cross-sections for \ce{Py-B(OH)3} obtained (i) from the full set of 500 sampled geometries using ADC(2)/cc-pVDZ (dashed blue line) and ADC(2)/aug-cc-pVDZ (solid blue line) and (ii) from the subset of 403 sampled geometries exhibiting a \ce{B-N} bond length shorter than 2.1 \AA~using ADC(2)/cc-pVDZ (dashed black line) and ADC(2)/aug-cc-pVDZ (solid black line).} 
\end{figure}

\begin{figure}
    \centering
    \includegraphics[width=1.1\linewidth]{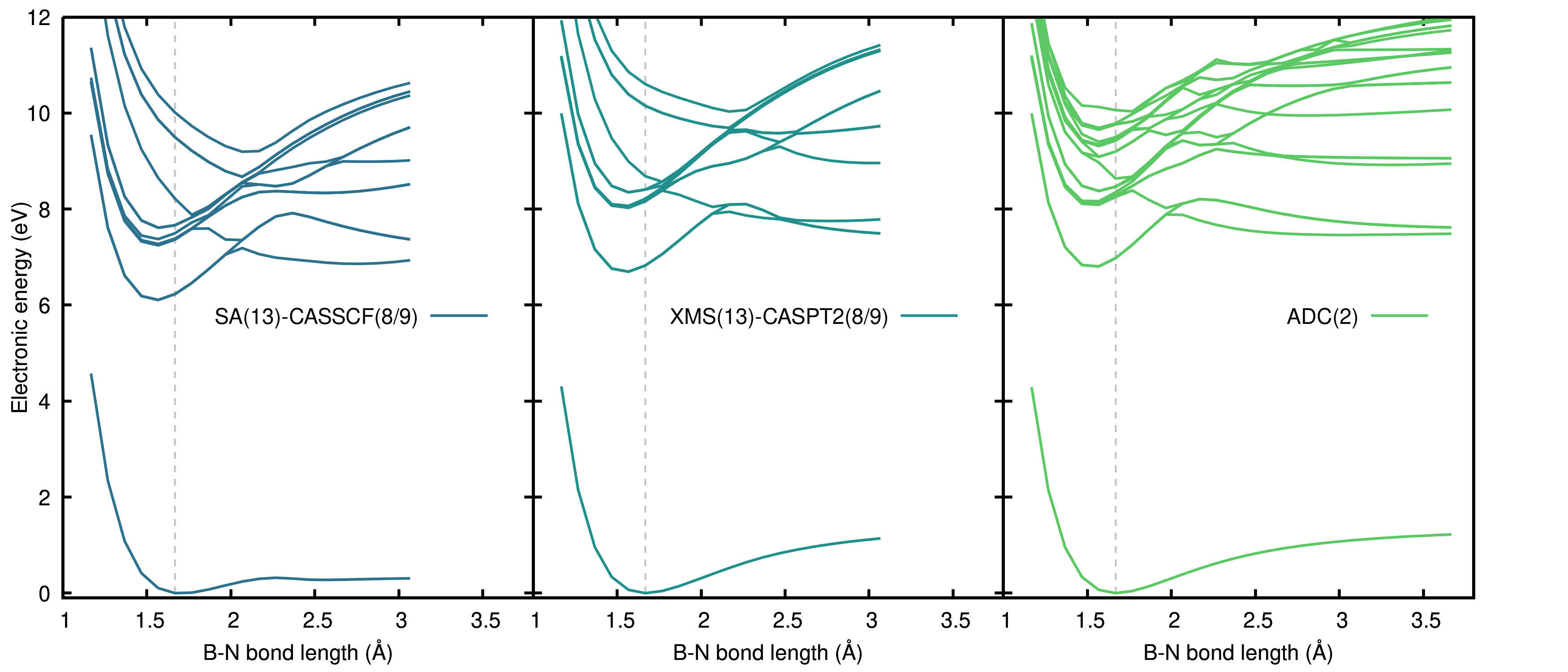}
    \caption{Electronic energies obtained for the ground-state relaxed scan (MP2/aug-cc-pVDZ) along the \ce{B-N} bond length with SA(13)-CASSCF(8/9)/aug-cc-pVDZ (dark blue, left panel), XMS(13)-CASPT2(8/9)/aug-cc-pVDZ (blue-green, middle panel), and ADC(2)/aug-cc-pVDZ (green, right panel). The last points of the relaxed scan (long \ce{B-N} bond lengths) are not reported for SA(13)-CASSCF(8/9)/aug-cc-pVDZ and XMS(13)-CASPT2(8/9)/aug-cc-pVDZ due to convergence issues. The location of the FC point (MP2) is indicated by a gray vertical dashed line.}
\end{figure}

\begin{figure}
    \centering
    \includegraphics[width=\linewidth]{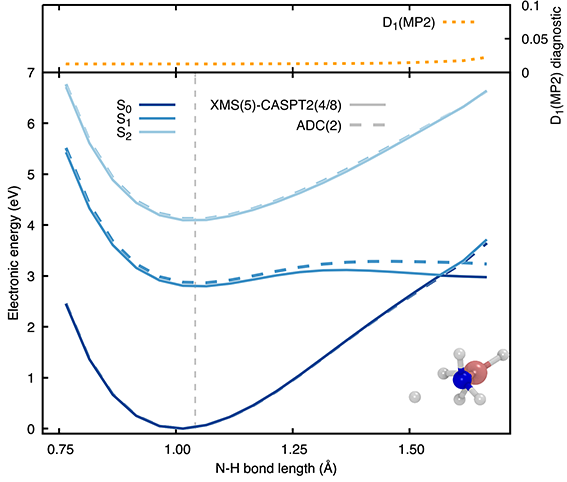}
    \caption{Relaxed scan along the \ce{N-H} bond of \ce{H3N-BH3} for the the S$_1$ excited electronic state obtained with ADC(2)/aug-cc-pVDZ. The \ce{N-H} bond length at the S$_1$ minimum-energy geometry is indicated by the gray vertical dashed line. The three lowest excited electronic states are depicted from dark (S$_0$) to light (S$_2$) blue, while the ADC(2)/aug-cc-pVDZ and XMS(5)-CASPT2(4/8)/aug-cc-pVDZ energies are indicated with dashed and solid lines, respectively. The D$_1$ diagnostic for the MP2 ground state is reported as a dotted orange line in the upper panel. The last structure of the relaxed scan is given as an inset.}
\end{figure}